\PassOptionsToPackage{table,xcdraw}{xcolor}

\documentclass[conference,compsoc]{IEEEtran}
\usepackage{graphicx} 

\usepackage[hidelinks]{hyperref}
\usepackage{amsmath,amssymb,amsfonts}
\usepackage{graphicx}
\usepackage{xcolor}
\usepackage{multirow}
\usepackage{rotating}
\usepackage{tablefootnote}
\usepackage{tabularx}
\usepackage{tikz}
\usepackage[english]{babel}
\usepackage[utf8]{inputenc}
\usepackage{lipsum}
\usepackage{listings}
\usepackage{tabularx}
\usepackage{siunitx}
\usepackage[nolist]{acronym}
\usepackage{booktabs}
\usepackage{comment}
\usepackage{xspace}
\usepackage{tablefootnote}
\usepackage{url}
\usepackage{makecell}

\def\orcid#1{\kern.08em\href{https://orcid.org/#1}{\protect\includegraphics[keepaspectratio,width=0.7em]{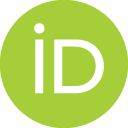}}}

\title{Digital Security --- A Question of Perspective\\A Large-Scale Telephone Survey with Four At-Risk User Groups}

\author{

\IEEEauthorblockN{
Franziska Herbert\IEEEauthorrefmark{1}\orcid{0000-0002-0842-9632},
Steffen Becker\IEEEauthorrefmark{1}\IEEEauthorrefmark{2}\orcid{0000-0001-7526-5597},
Annalina Buckmann\IEEEauthorrefmark{1}\orcid{0000-0002-7959-9743},
Marvin Kowalewski\IEEEauthorrefmark{1}\orcid{0000-0002-6929-3679}, \\
Jonas Hielscher\IEEEauthorrefmark{1}\orcid{0000-0002-5159-3868}, 
Yasemin Acar\IEEEauthorrefmark{3}\orcid{0000-0001-7167-7383},
Markus Dürmuth\IEEEauthorrefmark{4}\orcid{0000-0001-5048-3723},
Yixin Zou\IEEEauthorrefmark{2}\orcid{0000-0002-9088-705X}, and
M. Angela Sasse\IEEEauthorrefmark{1}\orcid{0000-0003-1823-5505}
}

\IEEEauthorblockA{
\IEEEauthorrefmark{1}Ruhr University Bochum, Germany \IEEEauthorrefmark{2}Max Planck Institute for Security and Privacy, Germany \\ \IEEEauthorrefmark{3}Paderborn University, Germany
\IEEEauthorrefmark{4}Hannover University, Germany
}
Email: \{\href{mailto:franziska.herbert@rub.de}{franziska.herbert}, \href{mailto:steffen.becker@rub.de}{steffen.becker}, \href{mailto:annalina.buckmann@rub.de}{annalina.buckmann}, \href{mailto:marvin.kowalewski@rub.de}{marvin.kowalewski}, \href{mailto:jonas.hielscher@rub.de}{jonas.hielscher}, \href{mailto:angela.sasse@rub.de}{angela.sasse}\}@rub.de, \\

\href{mailto:yasemin.acar@uni-paderborn.de}{yasemin.acar@uni-paderborn.de}, \href{mailto:markus.duermuth@itsec.uni-hannover.de}{markus.duermuth@itsec.uni-hannover.de}, \href{mailto:yixin.zou@mpi-sp.org}{yixin.zou@mpi-sp.org}

}

\newcommand{\ie}{i.\,e.}
\newcommand{\eg}{e.\,g.}

\newcommand{\etal}{et~al.\@\,}

\newcommand{\todo}[1]{}
\newcommand{\rev}[1]{{\color{black} #1}}

\newcommand{\umbrellaGroupCharacterization}{at-risk groups\xspace} 
\newcommand{\umbrellaSurveyConstructs}{security experiences\xspace} 

\begin{document}

\maketitle
\thispagestyle{plain}
\pagestyle{plain}

\begin{abstract}
This paper investigates the digital security experiences of four at-risk user groups in Germany, including older adults (70+), teenagers (14-17), people with migration background\rev{s}, and people with low formal education. 
Using computer-assisted telephone interviews, we sampled 250 participants per group, representative of region, gender, and partly age distributions. 
We examine their device usage, concerns, prior negative incidents, perceptions of potential attackers, and information sources.
Our study provides the first quantitative and nationally representative insights into the digital security experiences of these four at-risk groups in Germany. 
Our findings \rev{show} that participants with migration background\rev{s} used the most devices, sought more security information, and reported more experiences with cybercrime incidents than other groups.
Older adults used the \rev{fewest} devices and were least affected by cybercrimes. 
All groups relied on friends and family and online news as their primary sources of security information, with little concern about their social circles being potential attackers.
\rev{We} highlight the nuanced differences between the four at-risk groups and compare them to the broader German population when possible. 
We conclude by presenting recommendations for education, policy, and future research aimed at addressing the digital security needs of these at-risk user groups.
\end{abstract}
\section{Introduction}
\label{sec:intro}

Recent advancements in usable security and privacy research have challenged the notion of a ``general user'' \cite{kaur2021human,renaud2022accessible}, argued to acknowledge user diversity~\cite{fritsch2010towards}, and shown the specific threat models, risks, and needs of various \umbrellaGroupCharacterization~\cite{sannon2022privacy,mcdonald2020politics,walker2019moving,wang2017third} -- individuals and communities with ``risk factors that augment or amplify their chances of being digitally attacked and/or suffering disproportionate harms''~\cite{warford2022sok}. At-risk populations generally face disproportionate challenges in protecting their security and privacy due to disparities in digital literacy, resources, time, and linguistic and cultural barriers~\cite{sannon2022privacy,warford2022sok}.
Including \umbrellaGroupCharacterization in research helps generate insights on their unique risks and needs, and inform more inclusive design of technologies and educational materials.





While there has been a growing body of research on \umbrellaGroupCharacterization, such research is predominantly qualitative \cite{sannon2022privacy} and conducted with small and specific groups, such as the LQBTQ+-community \cite{scheuerman2018safe, blackwell2016lgbt, lerner2020privacy}, undocumented migrants or refugees \cite{guberek2018keeping, simko2018computer, coles2019accessing}, and survivors of intimate partner abuse \cite{freed2019my, havron2019clinical, slupska2021threat, matthews2017stories},
leading to valuable population-specific findings. Warford~\etal took the first step of synthesizing contextual risk factors across different at-risk populations, emphasizing societal factors (\eg, legal or political), relationships (\eg, reliance on a third party), and personal circumstances (\eg, constrained resources, lack of accessibility)~\cite{warford2022sok}. However, to our knowledge, there has not been any large-scale study with representative samples for these groups, which enables quantitative comparisons to identify the similarities and differences between multiple \umbrellaGroupCharacterization regarding their \umbrellaSurveyConstructs.\footnote{\rev{We use the term ``security experiences'' to refer to participants' device usage, security and privacy concerns, previous negative security- and privacy-related incidents, perceptions of potential attackers, and security information sources. The topics represent core elements of participants' digital security experiences and relevant context.}}

Our study \rev{makes the first} step to addressing this gap with a large-scale computer-assisted telephone interview (CATI) study  with $1003$ participants in total from four at-risk groups in Germany: older adults (70+), teenagers (14-17),\footnote{For participating in a survey in Germany, the ability to consent is crucial. As German law views teenagers from 14 years onward as criminally responsible (see §19 StGB, §§1 Abs. II, 3 JGG) consent ability is assumed.} people with migration background\rev{s},\footnote{``Migration background\rev{s}'' is a specific category mostly used in German-speaking countries, referring to residents who either have at least one parent who was born outside of Germany, who themselves migrated to Germany, and/or who hold a foreign citizenship~\cite{eu-06-migration}.} and people with little formal education (less than high school).\footnote{International Standard Classification of Education~(ISCED)~0-2~\cite{unesco-12-isced}} \rev{Past literature has characterized the} four groups as \umbrellaGroupCharacterization due to various reasons, \eg, teenagers are especially at risk of experiencing cyberbullying~\cite{Quayyum.2021} and older adults are especially at risk of \rev{being targeted for} tech scams and romance fraud~\cite{FBI.ElderFraudReport.2020}.
The four groups also have varying degrees of representation in existing usable security and privacy literature, 
\eg, there is much more related work on older adults than on people with low formal education, \rev{but all groups have been less represented in large-scale quantitative studies~\cite{redmiles-19-mturk-generalization, tang-22-generalize}. 
We focus on these four groups} considering the quantitative nature of our research, \rev{as} each group \rev{is} broad enough to gather a nationally representative sample of \rev{each groups'} 
population in terms of gender, region, and partly age. 
 
Our telephone survey collected participants' responses on the following topics: device usage, concerns, prior negative incidents, perceptions of potential attackers, and information sources. Knowing which devices are being used helps contextualize the risks each group might encounter. Eliciting their security concerns, potential attackers, and prior negative incidents helps understand the specific threat landscape for each population. 
The experience of threats and concerns may motivate information-seeking behavior, and identifying the information channels they consult (or not) informs how to target \rev{each} groups. 

Our research is guided by the following questions:

\begin{itemize}
    \item[\textbf{RQ1:}] What are the security experiences (device usage, concerns, prior negative incidents, perceptions of potential attackers, information sources) for older adults, teenagers, people with migration background\rev{s}, and people with low formal education? 
    \item[\textbf{RQ2:}] How similar or different are the four groups in terms of their security experiences?
\end{itemize}

\textbf{Summary of key findings.} 
Participants with migration background\rev{s} stand out for using the most devices, seeking more security information than the other groups, but also reporting most experiences with cybercrime incidents. 
In contrast, older adults used \rev{the fewest} devices and were least affected by cybercrime. 
Participants shared diverse concerns about digital security and there was no predominant concern across all groups. 
In terms of similarities, many participants across the groups reported being affected by malware and turning to their inner social circle (\ie, family and friends) for information on digital security, \rev{without} considering them as potential attackers. 
We compare our findings to prior (mostly qualitative) work and the broader German population when possible. 
For example, we find that our participants encountered cybercrime incidents much more frequently than the average German population~\cite{Digitalbarometer2022}, providing empirical evidence that these groups are indeed at higher risks of cybercrime and underscoring the need to protect these groups through \rev{better} educational efforts and \rev{public policy}.

\section{Group-Specific Prior Research}
\label{sub:relatedWork:FourGroups}
\rev{We} summarize the key findings of related work regarding the four at-risk groups. 
While our four sample groups tend to be underrepresented in quantitative security and privacy research~\cite{redmiles-19-mturk-generalization, tang-22-generalize}, several (mostly qualitative) studies have investigated these at-risk groups.


\subsection{Older Adults} 
There has been a considerable body of qualitative research on older adults' perceptions of privacy and security, both broadly~\cite{frik2019privacy,hornung2017navigating,Ray2019} and with regard to specific contexts such as social media~\cite{luders2017my,quan2018revisiting} and healthcare~\cite{demiris2008findings,boise2013willingness}. For example, Frik~\etal's study highlights older adults' differing attitudes toward privacy versus security, misconceptions about data flows, and blind spots in mitigation strategies, \rev{making them limit or avoid} technology use altogether~\cite{frik2019privacy}. 

Similarly, Ray~\etal's study shows how the perceived vulnerability of private information leaves many older adults anxious or frustrated, causing them to shy away from using online services~\cite{Ray2019}. Older adults also have particular needs that have not been sufficiently considered in mainstream security and privacy mechanisms. For example, older adults may find it challenging to manage passwords due to memory difficulties~\cite{hornung2017navigating,ray2021older}, and because of that the management is often delegated to friends or family members. 

Regarding advice sources, older adults were found to value social resources over expert advice, and they avoid using the Internet for cybersecurity information despite using it \rev{for} other \rev{topics}~\cite{nicholson-19-headline}. 


\subsection{Teenagers}
Teenagers tend to be digital natives as they are increasingly using digital technology from an early age~\cite{choong_2019}. The easy and nearly constant access to the internet comes with risks, not only for typical cybersecurity incidents (\eg, phishing, hacking, and identity theft)~\cite{Quayyum.2021} but also for events that can threaten one's psychological or even physical safety (\eg, exposure to unwanted explicit content, harassment, and sexual solicitations)~\cite{Mitchell_2014}.

Prior work has found that teens ``make online disclosures and render themselves more susceptible to experiences of risky online interactions''; which in turn generate privacy concerns, advice-seeking, and risk-coping behaviors~\cite{jia2015risk}. Resilience is not only a key factor protecting teens from experiencing online risks, but also neutralizes negative psychological effects associated with Internet addiction and online risk exposure~\cite{wisniewski2015resilience}. In terms of information sources, teenagers often turn to peers and online platforms to seek support on topics like online sexual interactions~\cite{razi2020let}, but are more reserved in discussing risky experiences with parents~\cite{wisniewski2017parents}. Also, parent involvement \rev{through} control (\ie, control apps) was associated with \rev{increased} risks~\cite{ghosh2018matter}.  

\subsection{Migration Background\rev{s}}
There have been multiple studies about the digital experiences of people with migration background\rev{s} especially refugees~\cite{simko2018computer, coles2019accessing, wyche2012we, brown2014reflection, Stapf2019, Kutscher2015, Gouma2020, coles2018new, lingel2014city, guberek2018keeping}. Guberek et al.'s study with undocumented immigrants in the United States identifies key concerns about identity theft, privacy, and online harassment, but participants' concerns about government surveillance are vague and \rev{intertwined} with resignation~\cite{guberek2018keeping}. Similarly, Simko et al.'s study with refugees shows how reliance on technology (\eg, for finding jobs and establishing a life) forces security ``best practices'' into the background~\cite{simko2018computer}. Focusing on people who recently migrated to Germany, Stapf~\cite{Stapf2019} found that they \rev{are} familiar with concepts of misinformation and hate speech on social media, but also value social media for information seeking and counseling; meanwhile, information from official sources and websites \rev{is} perceived as inaccessible, hard to understand, and not always helpful compared to information shared in their own language or based on other's personal experiences.

\subsection{Low Formal Education}
There is less related work on people with little formal education compared to other groups, \rev{and most related work we find for this group has been conducted in the US context}. \rev{Among} research on education and its effect on one's security and privacy experiences, \rev{it} has been found that individuals with \rev{lower} education tend to be less concerned about online privacy issues~\cite{sheehan2002toward} \rev{and doubt their Internet service providers' ability to protect their personal information~\cite{madden2017privacy}}, whereas those more educated are more likely to utilize privacy protection measures~\cite{rainie2013anonymity}, such as reading privacy policies~\cite{o2001analysis,milne2004strategies}. \rev{Bergström's study highlights the particular concerns held by} people with lower education regarding information search, email handling, and using debit cards~\cite{bergstroem2015concerns}. \rev{On the topic of viruses and hackers, Wash and Rader's study suggests that internet users with less education are more likely to show resignation; those with higher education report taking more protective actions but also rarely consider themselves to be vulnerable~\cite{wash2015too}.} 

\rev{Other studies have used lower socioeconomic status~(SES) as a proxy for lower education. 
A Pew 2017 survey highlights a knowledge gap on issues around privacy and security, as respondents with lower education scored lower in a 13-item quiz~\cite{olmstead2017americans}.} 
On the contrary, Redmiles~\etal's study finds that people's reported experiences of negative incidents are significantly related to advice sources, regardless of their SES or resources~\cite{redmiles_digital_Divide}.

\section{Research Method}
\label{Method}

To draw representative samples for the four at-risk groups, we employed computer-assisted telephone interviews~(CATIs). 
We chose CATIs for several reasons. 
(1)~CATIs allow for misunderstandings to be clarified as participants have the opportunity to ask questions. 
(2)~Compared with computer-assisted personal interviews~(CAPIs), which also \rev{allow clarifying} questions, CATIs are less expensive, and more participants can be interviewed in less time~\cite{yan2015computerassisted,Cervantes_2007_Methods}. 
(3)~For open-ended questions, participants are likely to provide more information in CATIs than in online surveys as they do not have to type out their answers, which also helps increase data quality. 
(4)~CATIs enable us to collect both qualitative and quantitative data, which can provide complementary insights. 
(5)~CATIs are generally considered a high-quality data collection method compared to paper and pencil surveys~\cite{yan2015computerassisted}.

CATIs also offer particular advantages for recruiting our target populations. 
Almost every household in Germany can be reached by telephone,\footnote{In 2022, 99.9\% of German households were equipped with either a landline or cell phone~\cite{destatis2022ICT}.} so the majority of the population -- including our four groups -- can be reached. 
Unlike online surveys, CATI participants do not need to be highly computer literate -- this is an important factor to consider as our target groups include older adults and people with low levels of formal education~\cite{yan2015computerassisted}. 
In fact, prior work has recommended using telephone surveys to reach populations such as older adults~\cite{Cervantes_2007_Methods}. 
Our data collection \rev{also} took place during the COVID-19 pandemic, which made in-person interviews difficult and CATIs a better alternative.

\subsection{Questionnaire}
Our CATI questionnaire contains fourteen closed questions and one open-ended question. 
We pilot-tested \rev{the questionnaire in multiple rounds}
to verify \rev{that} the duration is manageable and the questions are understandable over the phone. 
Below we provide an overview of all questions used for this work. 
The entire questionnaire can be found in \hyperref[appendix:interview]{Appendix~\ref{appendix:interview}}. 

\subsubsection{Introduction and Informed Consent}
Before starting with the actual questionnaire, interviewers introduced themselves, briefly stated the aim of the research and the planned duration of the interview.
Participants were informed that they could terminate the interview at any time, that the telephone interview would not cause any harm, and that they could \rev{choose} not \rev{to} respond to any question. 
We only interviewed participants after they provided informed consent. 

\subsubsection{Demographics and Technology Usage}
\label{method:demographics}
The first five questions were demographic screening questions to filter participants \rev{for} the desired target groups with the intended representativeness criteria.
If interviewees did not fit any of the at-risk groups, the interviewer politely ended the interview.
If interviewees were eligible, the interview began with a question about the device types they use in their daily lives~(Q1). We included the question since device usage may influence one's exposure to corresponding threats. 
For example, mobile devices face particular threats related to location tracking~\cite{Husted2010Mobile, Michalevsky2015PowerSpy, Givehchian2022Evaluating}. 
There are also particular vulnerabilities for IoT devices, such as weak voice authentication~\cite{lei2018insecurity} and \rev{continuous} listening and recording~\cite{jackson2018study}.


\subsubsection{Concerns, Prior Incidents, Advice Sources, and Potential Attackers}

Our next questions queried participants' concerns and prior experience with digital security threats. These questions help generate insights and implications for how to eliminate unnecessary fears and misconceptions among our participants and provide them with relevant information on protective strategies.

To avoid priming, this segment started with an open-ended question eliciting participants' concerns about their digital security~(Q4).
In a follow-up question, participants were asked whether they had already experienced various types of cybercrime identified in a survey by the German Federal Office for Information Security (Bundesamt für Sicherheit in der Informationstechnik; BSI)~\cite{zindler-20-digitalbarometer}, such as malware, phishing and cyberbullying~(Q6).
We then asked participants if and where they seek information about digital security~(Q7 and~Q8) to identify the channels best used to reach each group. 

To elicit participants' threat models and identify possible blind spots, we provided participants with eight groups of potential attackers (\eg, family members or officials from Germany) and asked them to rate how likely each group might pose a risk to their digital security (\eg, by obtaining unauthorized access to their personal data, stalking them online, or restricting their access to digital services)~(Q10). 
Response options consisted of a five-point rating scale ranging from \textit{1--not likely} to \textit{5--very likely}.

\subsection{CATI Implementation and Panels}

\def\rot{\rotatebox{90}}
\begin{table*}[t]
\small
\caption{Demographics and device usage of the four at-risk groups.}
    \label{tab:demographics}

\tabcolsep=0.5cm
    \begin{tabular}{ll|S[table-format=3]S[table-format=3]|S[table-format=3]S[table-format=3]|S[table-format=3]S[table-format=2]|S[table-format=3]S[table-format=3]}
    \toprule
                                                &   & \multicolumn{2}{c|}{\textbf{Older Adults}}            & \multicolumn{2}{c|}{\textbf{Teenagers}}            & \multicolumn{2}{c|}{\textbf{Migra. Backgr.}}         & \multicolumn{2}{c}{\textbf{Low Education}} \\
                                                &   & \multicolumn{2}{c|}{n=250} & \multicolumn{2}{c|}{n=250}  & \multicolumn{2}{c|}{n=251} & \multicolumn{2}{c}{n=252} \\
                                                &   & \multicolumn{1}{c}{n}      & \multicolumn{1}{c|}{\%}             & \multicolumn{1}{c}{n}   & \multicolumn{1}{c|}{\%}             & \multicolumn{1}{c}{n}   & \multicolumn{1}{c|}{\%} & \multicolumn{1}{c}{n}   & \multicolumn{1}{c}{\%} \\

                                                \cline{1-6} \hline
    \multirow{4}{*}{\rot{\textbf{Age}}}         & \multicolumn{1}{l|}{14-17}            & 0 & 0 & 250   & 100   & 0  & 0 & 0  & 0 \\
                                                & \multicolumn{1}{l|}{17-35}            & 0   & 0   & 0   & 0   & 103  & 41 & 39  & 15 \\
                                                & \multicolumn{1}{l|}{36-50}            & 0   & 0   & 0   & 0   & 63  & 25 & 48  & 19 \\
                                                & \multicolumn{1}{l|}{51-65}              & 0   & 0   & 0 & 0 & 57  & 23 & 68 & 27 \\
                                                & \multicolumn{1}{l|}{66-69}              & 0   & 0   & 0 & 0 & 14  & 5 & 75 & 30 \\
                                                & \multicolumn{1}{l|}{70+}              & 250   & 100   & 0 & 0 & 14  & 5 & 22 & 9 \\
                                                \cline{1-6} \hline
    \multirow{3}{*}{\rot{\textbf{Gender}}}      & \multicolumn{1}{l|}{Male}             & 130 & 52& 111 & 44             & 128 & 49 & 128 & 51 \\
                                                & \multicolumn{1}{l|}{Female}          & 120 & 48  & 138 & 55            & 123 & 51 & 124 & 49 \\
                                                & \multicolumn{1}{l|}{Non-binary}     & 0   & 0   & 1   & 1             & 0   & 0  & 0   & 0 \\ \cline{1-6} \hline
    \multirow{4}{*}{\rot{\textbf{Region}}}      & \multicolumn{1}{l|}{North}           & 39  & 16   & 42  & 17           & 37  & 15 & 38  & 15 \\
                                                & \multicolumn{1}{l|}{East}          & 45  & 18    & 51  & 20            & 28  & 11 & 24  & 10 \\
                                                & \multicolumn{1}{l|}{South}          & 76  & 30    & 70  & 28           & 82  & 33 & 86  & 34 \\
                                                & \multicolumn{1}{l|}{West}          & 90  & 36    & 87  & 35            & 104 & 41 & 104 & 41 \\ \hline
    \multirow{3}{*}{\rot{\textbf{Education}}}   & Low (ISCED 0-2)                    & 126 & 50   & 31  & 13                                              & 50  & 20                      & 252 & 100 \\
                                                & Medium (ISCED 3-4)                  & 57  & 22    & 63  & 25                                            & 103 & 41                      & 0   & 0 \\
                                                & High (ISCED 4-8)                    & 56  & 22   & 0   & 0                                              & 96  & 38                      & 0   & 0 \\
                                                & Other                              & 11  & 4   & 156\protect\footnotemark & 62                                               & 3   & 1                       & 0   & 0 \\
                                                \cline{1-6} \hline
    \multirow{6}{*}{\rot{\textbf{Device Usage}}} & \multicolumn{1}{l|}{Smartphone}      & 185 & 74 & 249 & 99    & 244 & 97 & 229 & 91 \\
                                                & \multicolumn{1}{l|}{Laptop or Desktop PC}        & 194 & 78   & 207 & 83    & 222 & 88 & 215 & 85 \\
                                                & \multicolumn{1}{l|}{Tablet}          & 88  & 35 & 138 & 55    & 144 & 57 & 121 & 48 \\
                                                & \multicolumn{1}{l|}{Smart Speaker}  & 23  & 9  & 54  & 21     & 72  & 29 & 46  & 18 \\
                                                & \multicolumn{1}{l|}{Wearables}        & 28  & 11 & 64  & 26   & 96  & 38 & 64  & 25 \\
                                                & \multicolumn{1}{l|}{None}       & 12  & 5 & 18  & 7    & 13  & 5 & 17  & 7 \\

    \bottomrule
    \end{tabular}
    
    \end{table*}
    %
\footnotetext{Many participants in this group had not yet graduated from school.} 
Interviews were conducted between October and December 2021 by professional telephone interviewers. 

Phone number sampling was based on a master sample, which contained up-to-date information on the range of numbers available in the German telephone network, with a distribution of $70\%$ landline numbers and $30\%$ mobile numbers generated at random~\cite{callegaro2011combining, cox1987constructive}. 
Prior to data collection, we conducted a training session with the interviewers, explaining the purpose of the study and guiding them step-by-step through the questionnaire, thereby giving them the opportunity to ask questions. 
Additionally, we supplied the interviewers with a glossary to prepare them for any potential questions from participants.
During the interview, the interviewers introduced themselves on behalf of our institution. 
If the invitee agreed to participate, their responses were recorded by interviewers using a web interface. 
The telephone interviews were conducted on multiple days of the week and at several times of the day. 
Participants who terminated the interview early or withdrew consent to analyze their data were excluded from the final data set.
\rev{We cannot make any statements about the response rates because the CATI provider did not disclose them to us.}

For each of the four groups, about $250$ participants were interviewed, resulting in a total of $1003$ telephone interviews (see \autoref{tab:demographics} for further demographic details on each group). 
All four groups were sampled to be representative of \rev{their respective gender and regional distribution} in Germany~\cite{destatis2019mikrozensus, destatis2020fortschreibung}. 
The groups with low formal education and migration background\rev{s} were \rev{each} sampled to be \rev{group-}representative in terms of age as well. 
\rev{
For example, the quota for participants with migration backgrounds aged 35 to 54 was 95 participants.} Our samples matched the target quotas with only small discrepancies ranging from $1\%$ to $4\%$ with the exception of gender quotas for teenagers and older adults, which were met with $8$ respectively $9\%$ deviation.
As common with CATIs, participants were not compensated. 

\rev{There are natural overlaps between the four groups as a result of our sampling strategy (e.g., some older participants in the older adults group may also have migration backgrounds and/or low levels of formal education).
These overlaps reflect the representative composition of these groups in Germany and allow us to display the diversity of these groups as it exists.
}

\subsection{Data Analysis and Coding Procedure}
\label{section:dataanalysis}
All data was first analyzed descriptively, except for the open-ended question~(Q4). 
We then performed $X^2$ tests for all differences greater than $5\%$ to determine which groups differ significantly regarding binary data~(Q1, Q6, Q7).
We used Bonfferoni-Holm alpha correction to account for multiple testing. 
To identify significant differences between the groups for interval data~(Q9), we used (Welch-)ANOVAs with Bonferroni corrected post-hoc tests. 
Regarding effect sizes, we report $phi$ for $X^2$ tests and \textit{Cohen's d} for the ANOVA post-hoc tests~\cite{cohen2016power}. 
 
For analyzing text responses (Q4), we used qualitative content analysis~\cite{mayring_qualitative_2014}. 
Three researchers coded the first $150$ responses independently, compared their codes, and agreed upon an initial codebook. 
Subsequently, the same three researchers coded the remaining responses independently.
We calculated Fleiss' Kappa to determine inter-rater reliability~(IRR), which was considered moderately acceptable ($\kappa=0.74$)~\cite{McDonald_2019}. 
Finally, the three researchers jointly reviewed unclear cases and summarized the codes under broader categories. 
We provide the complete codebook and code distributions for each group in 
\hyperref[appendix:codebookCATI]{Appendix~\ref{appendix:codebookCATI}}.

\subsection{Positionality Statement}
As researchers, we are aware that our own backgrounds, values, and biases influence how we conduct research~\cite{liang2021embracing} and we are always at risk of reproducing knowledge that reifies power~\cite{sweet2020knows}. 
We strive to accurately represent the perspectives of the groups we study while critically reflecting on our approach. 
Our team comprises highly-educated researchers from the disciplines of psychology, usable security and privacy, and security engineering. 
None of us belong to the groups of teenagers, older adults, or people with lower formal education. 
We acknowledge that our relative privilege within society (particularly high education) provides us with certain advantages that our participants do not hold. 
However, two members have conducted research with older adults in the past, and two members have migration background\rev{s} themselves, which gives us relevant insights into these groups.

\subsection{Ethics} 
As our department does not have an institutional review board, we had extensive ethics-related discussions within our interdisciplinary team. 
We also developed a protocol that followed best practices of human subjects research~\cite{dhs-12-menlo-report} and data protection guidelines, including the European General Data Protection Regulation (GDPR). All data protection measures were reviewed and approved by our institution's data protection office. Additionally, the CATI provider signed an agreement with our institution to follow GDPR guidelines.
We ensured accessible language to not overwhelm participants or leave them frightened after the interview. We also followed current German law allowing teenagers of 14 years and above to take part in surveys without their parents' consent.
The CATI research method and our provider did not allow us to compensate our participants financially or to provide them with information on digital security. However, we hope our publications will provide useful insights
for our target groups so they may benefit in the long run.

\subsection{Limitations}
First, our study was conducted with German residents only, and our findings might not generalize to other countries or societies. 
Second, some questions required participants to admit gaps in their knowledge or mistakes they may have made -- something they may be less likely to answer truthfully compared to neutral questions~\cite{redmiles-19-mturk-generalization}. We tried to overcome this limitation through careful questionnaire design and by letting participants know that there are no wrong answers and they would not be judged. 
Third, to avoid bias, especially for the open-ended question~(Q4), we decided against randomizing the question order. We thus can not preclude response order effects.
Fourth, there are other at-risk groups that should be represented in research~\cite{renaud2022accessible}; for example, our provider did not offer the possibility to interview participants that are differently abled or children under 14. 
\rev{Fifth,} some of the related work referenced throughout the paper was performed before the COVID-19 pandemic. 
As the global population~\cite{nguyen2020covid19} as well as specific groups such as older adults~\cite{nimrod2020changes} experience changes in Internet usage during the pandemic, this has implications for their security experiences and needs to be considered when comparing our findings with previous studies. 
\rev{Finally, based on our research design, we can only describe group-level differences but cannot accurately identify specific reasons that explain these differences.}
\section{Results}
\label{Results}
Our results section is organized as follows: 
First, we present findings on device usage, before we move on to participants' concerns regarding their digital security and their past experiences with cybercrime. 
Finally, we report our findings on participants' perceptions of potential attackers and end with the results on the sources of information they use most frequently.

\subsection{Device Usage}
\label{DeviceUsage}

\autoref{tab:demographics} shows an overview of used devices per group:
Overall, all groups reported high use of smartphones and PCs, followed by tablets; IoT devices such as smart speakers and wearables were used less frequently.
Only 5 to 7 percent of each user group do not use any of the devices queried, \ie, the use of at least one Internet-enabled device is consistently high in all groups.
The average number of used device types is $3.0$ for teenagers, $2.3$ for older adults, $3.4$ for participants with migration background\rev{s}, and $3.0$ for participants with low formal education. Comparing our findings to the Germany-wide BSI survey~\cite{Digitalbarometer2022}, our participants' usage rates are higher for smartphones (89\%) and Laptops (71\%) but more similar to the German public for tablets (53\%) and wearables (22\%).

In terms of between-group differences, older adults had significantly lower device usage overall: their smartphone and wearable usage is significantly lower than that of the other three groups; they also use tablets and smart speakers less often than teenagers and participants with migration background\rev{s}. 
In addition, people with low formal education reported significantly less prevalent smartphone usage than teenagers and people with migration background\rev{s}. 
Participants with a migration background\rev{s} reported significantly higher rates of wearable use compared to the other groups.
Effect sizes for significant differences range between small ($0.13$) and moderate ($0.38$).

\begin{table*}[t]
\small
\caption{Digital security concerns~(Q4) among older adults, teenagers, people with migration background\rev{s}, and people with low formal education. The percentages relate only to those participants that report concerns. }
    \label{tab:q4concerns}

\tabcolsep=0.5cm
\renewcommand{\arraystretch}{1.2}
    \begin{tabular}{l|c|c|c|c}
    \toprule
    \textbf{Code}
    & \textbf{Older Adults}
    & \textbf{Teenagers}
    & \textbf{Migra. Backgr.}
    & \textbf{Low Education} \\
    
    & \% & \%  & \% & \% \\
    \midrule
    \textbf{Active attack} &&&&\\
    Hacker attack & 16 & 24 & 19 & 19 \\
    Financial loss & 21 & 7 & 15 & 12 \\
    Data theft & 6 & 14 & 19 & 11 \\
    Malware & 4 & 21 & 7 & 7 \\
    Password theft & 1 & 17 & 6 & 5 \\
    Phishing & 4 & 7 & 7 & 4 \\
    \midrule
    \textbf{Tracking} &&&&\\
    Data collection, aggregation, and use & 6 & 4 & 9 & 5 \\
    \midrule
    \textbf{Passive attack} &&&&\\
    Surveillance & 11 & 1 & 4 & 8 \\
    
    \bottomrule
    \end{tabular}
\end{table*}

\paragraph{\textbf{Summary}} Participants with migration background\rev{s} are the most active users of Internet-enabled devices, while older participants use them least commonly.
Teenagers and participants with low formal education are in the middle of the spectrum and show relatively small differences between each other.

\begin{figure*}
    \centering
    \includegraphics[width=1\linewidth]{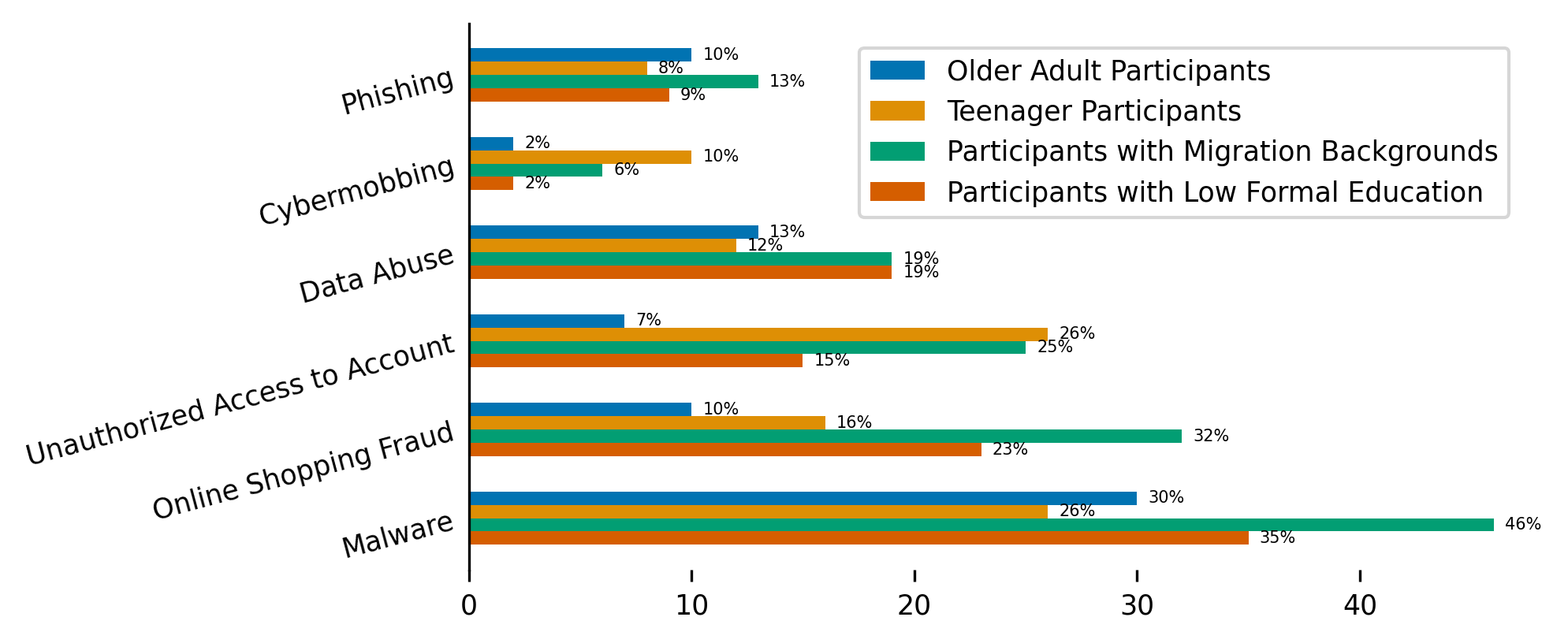}
    \caption{\rev{Participants from each group who were affected by different cybercrimes~(Q6), rounded to full percentages.}}
    \label{fig:Q6cybercrimes}
\end{figure*}

\subsection{Digital Security Concerns}
\label{results-Q4}
For the open question asking for participants' concerns regarding their digital security (Q4), the response rates are higher for people with migration background\rev{s} ($49\%$) and low formal education ($49\%$). In contrast, $38\%$ of teenagers and $34\%$ of older adults stated any digital security concerns. 
$12$ participants across all groups explicitly stated that they had no concerns (\eg, \textit{``I am not really concerned, because you have protections you can rely on. I make sure that I have the latest programs.''}).

The shared concerns varied significantly across all groups; no specific concern was raised by a majority of participants. This finding indicates that there is a wide diffusion of scattered risk awareness among these groups rather than a solidified ``body of knowledge.'' It could also be a reflection of users being overwhelmed by the large body of security advice that lacks prioritization~\cite{redmiles2020comprehensive, reeder-17-152-simple-steps}. 

We next present the more salient and prevalent concerns in~\autoref{tab:q4concerns} (including example quotes translated from German). The percentages stated in the following are based on only those who stated any concerns rather than all participants.

\paragraph{\textit{Hacking}} Attacks by ``hackers'' were one of the concerns named across all groups -- teenagers were slightly more concerned ($24\%$) than participants with migration background\rev{s} and those with low education (both $19\%$), older adults were slightly less concerned ($16\%$). One participant with migration background\rev{s} stated: \textit{``I am afraid of being hacked because I now do a lot of things and buy a lot of things online,''}. 

\paragraph{\textit{Financial loss}} The adult groups -- all groups except for the teenagers -- reported more concerns about financial loss compared to teenagers ($7\%$). Older adults had most concerns about financial loss ($21\%$), \rev{which might be linked with their transition to retirement and assets to manage at this life stage.}
In line with prior work~\cite{quan2020online}, older adults' concerns about financial loss were also intertwined with concerns about cybercrime such as scams: \eg, \textit{``Many scammers lurk on the internet and just want to rip off your money.''}

\paragraph{\textit{Malware and password theft}} Compared to the other groups, more teenagers reported concerns about malware (\eg, \textit{``I am concerned about viruses on the mobile phone and on my laptop''}) and password theft (\textit{``I am concerned that my passwords are hacked''}). Participants with migration background\rev{s} and participants with low education had these concerns to a similar extent, but these concerns were mentioned less frequently by older adults.

\paragraph{\textit{Data theft}} Participants with migration background\rev{s} were more concerned about data theft ($19\%$) than the other groups, especially compared to older adults ($6\%$). One participant with migration background\rev{s} put it this way: \textit{``I'm afraid that my data will be stolen without me knowing or wanting it.''} 
Older adults' lower concern \rev{about data theft could also be viewed together with our earlier findings about}
their relatively low device usage (\autoref{DeviceUsage}), \rev{which could imply} that they had less data ``out there.''

\paragraph{\textit{Phishing}} Phishing was mentioned infrequently across all groups (between $4\%$ and $7\%$), but the few participants that named phishing were aware of its risks and tried to take measures. One participant with low formal education stated: \textit{``In general I am afraid of spear phishing, as I have already been affected by it and I am always very cautious when opening unknown emails.''} Contrary to 
prior measurement studies that shows the prevalence of phishing attacks in the wild~\cite{Thomas_2017,han2016phisheye},~\rev{\cite{APWG_2022}} our participants seem not that concerned about phishing-related risks. 



\paragraph{\textit{Tracking and surveillance}} While concerns about data collection, aggregation, and use were also relatively low across the groups (between $4\%$ and $6\%$), participants with migration background\rev{s} stood out for having slightly higher concerns ($9\%$). \rev{This finding could be contextualized in prior work highlighting} the pervasive government surveillance they are experiencing~\cite{wang2022american} and \rev{our earlier finding about their higher device usage} 
(see~\autoref{DeviceUsage}). One participant commented on tracking from for-profit companies, who are also known to exchange data with government agencies~\cite{wang2022american}: \textit{``You never know how long or where data links are stored~\dots\ at Google Cloud, Whats App chat histories and online purchase service portals and apps~\dots\ They can keep track [of] consumers for a long time.''}

Concerns about surveillance were mentioned by more older adults ($11\%$) and participants with low formal education ($8\%$), especially compared to teenagers ($1\%$). These concerns often come with a sense of digital resignation~\cite{draper2019corporate}, \eg, \textit{``Everything is recorded, you can't hide anything''.}

\paragraph{\textbf{Summary}} Participants' concerns about digital security were diverse, and there was no dominant concern across all groups. The primary concern of older adults was about financial loss, whereas teenagers were most concerned about malware. Participants with migration background\rev{s} had the most concerns about data theft and tracking of all groups.

\subsection{Prior Experience with Cybercrime} 
\label{results-Q6}

We asked our participants if they had experienced any of the cybercrime incidents that were part of a German-wide online survey conducted by the BSI with $2000$ respondents aged 16-69~\cite{Digitalbarometer2022}. Across all groups, $55\%$ of our participants reported at least one incident -- indicating that the prevalence of cybercrime experiences in our sample is much higher than the $29\%$ in the BSI survey. 
\rev{The methodological differences between our work and the BSI survey should be noted when viewing this discrepancy --} the BSI survey was administered as a web survey, so participants might be more digitally literate and take more security measures. Our finding also validates the notion that at-risk groups are experiencing more cybercrime than ``average'' users.
 
The difference is especially pronounced for participants with migration background\rev{s} ($72\%$ reported experiences with at least one incident).  The percentage of prior experience with at least one cybercrime incident is comparatively lower for teenagers and participants with low formal education (both $54\%$) and for older adults ($42\%$). 

\paragraph{\textit{Malware}} A detailed examination of the specific types of cybercrime incidents (\autoref{fig:Q6cybercrimes}) shows that the most common incident was malware -- $30\%$ for older adults, $26\%$ for teenagers, $46\%$ for participants with migration background\rev{s}, and $35\%$ for participants with low formal education. Those numbers are much higher than in the BSI survey -- except for teenagers -- in which only $24\%$ reported being affected by malware~\cite{Digitalbarometer2022}. Participants with migration background\rev{s} experienced significantly more malware than teenagers ($X^2 = 19.63, p < 0.05$, $phi = 0.2$) and older adults ($X^2 = 12.65, p < 0.05$, $phi = 0.16$).
However, the effect sizes are only small. This finding also contrasts our earlier findings about concerns (\autoref{results-Q4}), as malware was rarely mentioned as a concern except for teenagers, indicating a misalignment between concerns and actual experiences.

\paragraph{\textit{Fraud in online shopping}}  Our participants' experiences with fraud in online shopping are generally in line with the BSI survey ($25\%$)~\cite{Digitalbarometer2022}. $32\%$ of participants with migration background\rev{s} reported experiencing this -- the percentage is significantly higher than that of older adults ($10\%$, $X^2 = 35.92, p < 0.05$) and teenagers ($16\%$, $X^2 = 18.21, p < 0.05$). The difference between older adults ($10\%$) and participants with low formal education ($23\%$) is also significant ($X^2 = 14.48, p < 0.05$). All effect sizes are rather small ($phi < 0.2$). Participants' experiences with fraud in online shopping are reflected in their concerns, as financial loss is a natural consequence of fraud and was mentioned as a concern by participants across the adult groups (\autoref{results-Q4}).

\paragraph{\textit{Unauthorized access to an online account}} Teenagers ($26\%$) and participants with migration background\rev{s} ($25\%$) reported most account compromises (unauthorized access to an online account). Older adults were the least affected ($7\%$), which is not surprising given their comparatively infrequent device usage (\autoref{DeviceUsage}). The differences are significant for teenagers versus older adults ($X^2 = 32.22, p < 0.05$, $phi = 0.26$) and teenagers versus participants with low formal education ($X^2 = 8.52, p < 0.05$, $phi = 0.14$). The differences between participants with migration background\rev{s} and older adults are also significant ($X^2 = 28.89, p < 0.05$, $phi = 0.25$), all with small effect sizes. The higher rate of teenagers being affected is in line with our findings about concerns, as $17\%$ of teenagers reported concerns about password theft which can directly lead to account compromises. 

\paragraph{\textit{Other cybercrimes}} For other less prevalent cybercrime incidents, all groups reported being equally affected by data abuse (percentages between $12\%$ to $19\%$). The lack of between-group differences also applies to phishing, which was experienced by fewer participants ($8\%$ to $13\%$) and corroborates the limited concerns about phishing in \autoref{results-Q4}. $10\%$ of teenagers reported experiences with cybermobbing, which is significantly more compared to older adults ($X^2 = 11.86, p < 0.05$, $phi = 0.16$) and participants with low formal education ($X^2 = 13.82, p < 0.05$, $phi = 0.17$), with small effect sizes. For ransomware, cyberstalking, and romance scam, the rates are lower than $5\%$ and similar across all groups; we did not detect any significant between-group differences.

\paragraph{\textbf{Summary}} Our participants encountered more cybercrime incidents compared to the average German population. Participants with migration background\rev{s} had the most negative incidents (particularly malware and fraud in online shopping), whereas older adults were less affected. Participants' actual experiences with cybercrime are generally in line with their concerns except for malware: teenagers were the most concerned but the least affected, and participants with migration background\rev{s} encountered significantly more malware than their concerns.

\subsection{Potential Attackers}
\label{results-Q9} 

\begin{table*}[tb]
\definecolor{tabcol}{RGB}{254,97,0}
\caption{Rated probability (mean values) of eight possible attacker groups posing a risk to the digital security of the participants. Rating scale ranging from 1--not likely to 5--very likely to pose risk.}
\centering
\begin{tabular}{lcccccccc@{\hspace{0.4cm}}}
\toprule
\textbf{Group} & \multicolumn{7}{c}{\textbf{Possible Attackers}} \\
\cmidrule{2-9}
& \makecell{Family \\ members} & Friends & \makecell{Work \\ colleagues} &  \makecell{ Officials from \\ Germany } & \makecell{Officials from \\ other countries} & \makecell{Private sector \\companies} & Criminals & Hackers \\
\midrule
Older Adults & \cellcolor{tabcol!32} $1.27$ & \cellcolor{tabcol!33} $1.31$ & \cellcolor{tabcol!33} $1.33$ & \cellcolor{tabcol!37} $1.48$ & \cellcolor{tabcol!67} $2.68$ & \cellcolor{tabcol!68} $2.71$ & \cellcolor{tabcol!85} $3.40$ & 
\cellcolor{tabcol!78} $3.13$ \\
Teenagers & \cellcolor{tabcol!38} $1.49$ & \cellcolor{tabcol!44} $1.74$ & \cellcolor{tabcol!39} $1.54$ & \cellcolor{tabcol!69} $2.76$ & \cellcolor{tabcol!69} $2.75$ & \cellcolor{tabcol!75} $3.00$ & \cellcolor{tabcol!97} $3.87$ & 
\cellcolor{tabcol!94} $3.74$ \\
Migra. Backgr. & \cellcolor{tabcol!32} $1.28$ & \cellcolor{tabcol!35} $1.40$ & \cellcolor{tabcol!42} $1.67$ & \cellcolor{tabcol!74} $2.95$ & \cellcolor{tabcol!82} $3.26$ & \cellcolor{tabcol!81} $3.22$ & \cellcolor{tabcol!99} $3.96$ & 
\cellcolor{tabcol!91} $3.63$ \\
Low Education & \cellcolor{tabcol!33} $1.33$ & \cellcolor{tabcol!37} $1.47$ & \cellcolor{tabcol!42} $1.67$ & \cellcolor{tabcol!75} $2.98$ & \cellcolor{tabcol!82} $3.26$ & \cellcolor{tabcol!83} $3.30$ & \cellcolor{tabcol!100} $4.00$ & \cellcolor{tabcol!90} $3.60$ \\
\bottomrule
\end{tabular}
\label{tab:Q10heatmap}
\end{table*}

Participants' perceptions of the risks different groups pose to their digital security (Q10) are displayed in~\autoref{tab:Q10heatmap}. 
Participants across all groups rarely perceived people close to them (\eg, family members, friends, and acquaintances) as possible attackers. Interestingly, teenagers identified friends and acquaintances as possible attackers significantly more than the other groups, although the average rating is still ``little likely'' ($M = 1.74$). The effect sizes were small, except for the post-hoc test for teenagers versus older adults, for which we observed a moderate effect size ($Cohen's~d = 0.53$). This finding also relates to our earlier finding about how teenagers reported more experiences with cybermobbing in~\autoref{results-Q6} and echoes past research on a high volume of cybermobbing incidents in school~\cite{alim2017cyberbullying}. 

Similarly, work colleagues were rarely perceived as possible attackers by teenagers, participants with migration background\rev{s}, and participants with low formal education ($M \approx 2$ for all three groups). Older adults were even less defensive against work colleagues ($M < 1.5$), and the differences between older adults and the other three groups are significant, with small effect sizes ($d < 0.5$). 

For officials from Germany (\eg, police, secret services, and the government), participants with migration background\rev{s} ($M = 2.95$) and participants with low formal education ($M = 2.98$) viewed them as possible attackers significantly more than older adults ($M = 2.68$). 
The same pattern also applies to officials from other countries and private sector companies. Officials from other countries were viewed significantly less as a risk by teenagers compared to participants with migration background\rev{s} and low formal education. The effect sizes are small ($d < 0.5$).

By contrast, criminals ``who want to get rich from your data'' were often identified as possible attackers by many participants. All groups except older adults thought they were ``quite likely'' to \rev{pose} risks, whereas older adults only viewed them as moderately risky; the differences between older adults and the other groups are all significant with small effect sizes. 
We found the same pattern for hackers ``who gain unauthorized access to data and devices for fun'' as older adults perceived them to be less of a threat than the other groups.

\paragraph{\textbf{Summary}} We do not observe mean values above $4$, showing that participants on average did not have high risk perceptions for the possible attackers we queried. Between the different groups, hackers and criminals were viewed as quite a threat, whereas those closer to participants such as family members, friends and acquaintances, and work colleagues were not. Across the four groups, older adults had significantly lower risk perceptions toward these possible attackers than the other groups: \rev{the lower risk perceptions could be contextualized in our finding about older adults' lower device usage and prior work on older adults being more trusting than younger adults~\cite{bailey2015trust}.}

\subsection{Information Sources}
\label{results-Q7}

To derive insights on how to best reach different groups for digital security education, we asked participants about their information sources, starting with a binary question on whether they actively seek information about digital security (Q7). Slightly above half of participants with migration background\rev{s} ($54\%$) reported doing this; the percentage is lower for older adults ($41\%$), participants with low formal education ($40\%$), and teenagers ($38\%$). The differences in information seeking between participants with migration background\rev{s} and the other groups are all significant, albeit with small effect sizes ($phi < 0.2$). Considering our previous finding that participants with migration background\rev{s} had disproportionately high encounters with cybercrime incidents, such experiences could serve as strong motivators and learning opportunities~\cite{zou2020examining}.

\begin{figure*}
    \centering
    \includegraphics[width=1\linewidth]{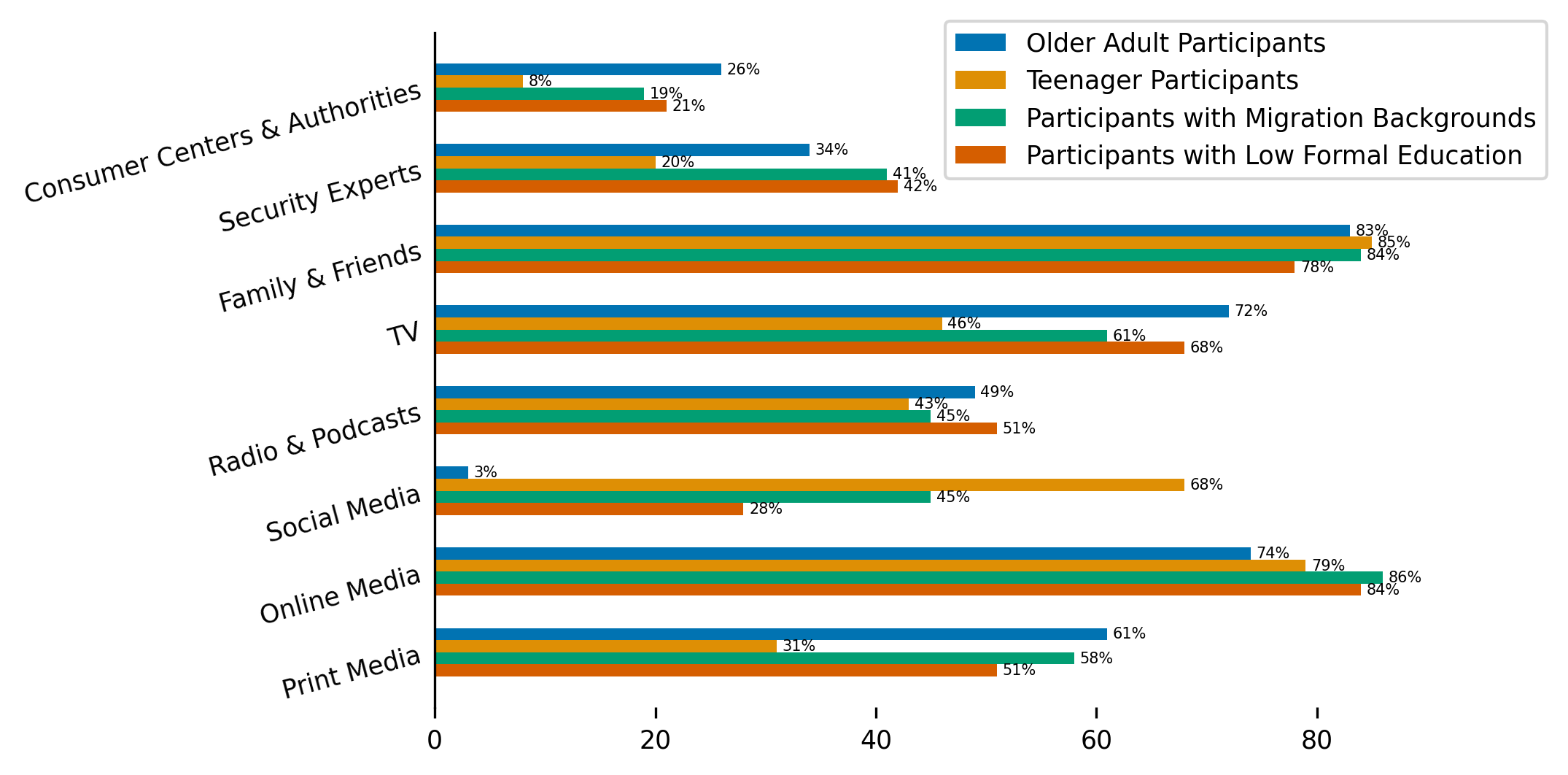}
    \caption{\rev{Participants' sources of information~(Q8), rounded to full percentages.} Only participants who answered ``yes'' to Q7 are included.}
    \label{fig:Q7securitysources}
\end{figure*}

To participants who said they seek information on digital security, we offered a list of possible sources (Q8). 
The most reported across all groups were friends/family (between $78\%$ and $83\%$) and online media (between $74\%$ and $84\%$) (see~\autoref{fig:Q7securitysources}), with no significant between-group differences.
 
About half of the participants across all groups obtained information from radio or podcasts (between $43\%$ and $51\%$), again with no significant between-group differences.

Teenagers consult print media much less frequently than other groups, which is also in line with teenagers' tech use patterns in general \cite{Vogels_2022, vomOrde_2023}. The differences are significant for the pairwise comparisons with older adults ($X^2 = 17.44, p < 0.05$, $phi = 0.31$) and with participants with migration background\rev{s} ($X^2 = 15.57, p < 0.05$, $phi = 0.27$), with small and moderate effect sizes. The same pattern also applies to TV, as teenagers relied on TV as a source of digital security significantly less than older adults ($X^2 = 12.34, p < 0.05$, $phi = 0.26$) and participants with low formal education ($X^2 = 8.50, p < 0.05$, $phi = 0.22$), with small effect sizes ($phi < 0.3$). 

In contrast, teenagers used social media significantly more than older adults ($X^2 = 91.20, p < 0.05$, $phi = 0.69$), participants with migration background\rev{s} ($X^2 = 11.23, p < 0.05$, $phi = 0.23$), and participants with low formal education ($X^2 = 30.31, p < 0.05$, $phi = 0.4$). Older adults use social media much less than the other groups, with also significant differences compared to participants with migration background\rev{s} ($X^2 = 50.98, p < 0.05$, $phi = 0.47$) and with participants with low formal education ($X^2 = 22.78, p < 0.05$, $phi = 0.35$). Almost all of these effect sizes are moderate ($phi > 0.3$). 

Lastly, teenagers also differ from other groups in their use of IT security experts and consumer advice centers/authorities as information sources. Teenagers used security experts significantly less than participants with migration background\rev{s} ($X^2 = 10.06, p < 0.05$, $phi = 0.22$) and participants with low formal education ($X^2 = 9.97, p < 0.05$, $phi = 0.24$). Teenagers also used consumer advice centers/authorities significantly less than older adults ($X^2 = 9.77, p < 0.05$, $phi = 0.24$).

\textbf{Summary.} While participants reported a variety of information sources, friends/family and online media were used more than others. The reliance on family and peers to \rev{navigate} digital security threats is also observed in SoKs of at-risk groups~\cite{sannon2022privacy,warford2022sok}. By contrast, our finding differs from Redmiles et al.'s study based on a US national representative sample~\cite{redmiles2016learned}, in which learning from friends/family was not as prevalent as from prompts (such as password meters and update reminders) and automated/forced security (\eg, automatic updates). In terms of between-group differences, we observe that participants with migration background\rev{s} were the most active information seekers across all groups. Teenagers also exhibited a unique pattern compared to the other three groups \rev{as they} rely more on social media and rely less on authoritative sources.

\section{Discussion}
\label{discussion}

We conducted a large-scale telephone survey with four at-risk groups -- older adults, teenagers, people with migration background\rev{s}, and people with low formal education -- using Germany-representative samples. 
This approach allows us to (1)~compare and contrast our findings with prior work (\autoref{subsec:compare-prior-work}), and (2)~systematically \rev{and consistently} compare the findings about digital security experiences \rev{across the four groups, who answered} the same questionnaires (\autoref{subsec:compare-between-group}). 
We conclude our paper by discussing how our findings guide future research as well as efforts for security education and policymaking (\autoref{subsec:recommendations}). 

\subsection{Comparison With Previous Studies (RQ1)}
\label{subsec:compare-prior-work}


In this section, we compare and contrast our study's findings with the (predominantly qualitative) prior research on each group presented in \autoref{sub:relatedWork:FourGroups}. 

\subsubsection{Older Adults}

Our findings add more nuances to prior work that frames older adults as an at-risk population~\cite{warford2022sok} or highlights the \textit{vulnerability} of older adults to various security and privacy risks~\cite{frik2019privacy}. 
In our study, 42\% of our older adult participants reported experiences with cybercrime. \rev{While} this rate is still much higher than the 29\% rate reported in the BSI survey~\cite{Digitalbarometer2022}, indicating that older adults are at higher risk of experiencing cybercrime than the general population in Germany, their rate is lower \rev{than} the other three groups. 
Combined with the finding that older adults use digital devices less frequently, this indicates that older adults may have a smaller attack surface in general.

Regarding information sources, our study revealed similar findings as those from Hornung~\etal~\cite{hornung2017navigating}: participants referred to friends and family for advice on how to protect their data. 
However, such reliance can introduce new risks when ``care surveillance''~\cite{essen2008two} and insider threats (\eg, financial abuse by a family member) happen. 

Nicholson~\etal~\cite{nicholson-19-headline} found that older adults value social resources over expert advice for cybersecurity information seeking; 
our older adult participants \rev{exhibited a similar pattern by seeking} information from friends and family, but they also used online media, TV, and print media \rev{as additional sources}, many of which could provide interfaces with experts.

Adding to prior research\rev{~\cite{frik2019privacy,quan2020online}}, we provide insights \rev{into} older adults' digital security concerns, such as financial loss and hacker attacks, and on their risk perception of different groups, which was only negligible to moderate for the queried groups (\eg, for family members or criminals).

\subsubsection{Teenagers}
 Our findings confirm that German teenagers are aware of some of the risks mentioned in Quayyum~\etal's study~\cite{Quayyum.2021}; hackers, malware, password, and data theft were the most commonly named concerns in our sample. The prevalence we found for teenagers being victimized by cybermobbing or cyberbullying is also in line with findings in related studies~\cite{Feierabend.2016,BARMER.2021}. On top of prior work that shows how teenagers engage in a broader variety of online activities than adults~\cite{dedkova2022digital}, including risky behavior such as installing questionable software~\cite{furnell2008security} (which can lead to malware), our findings further highlight the importance of supporting teenagers to cope with malware: our teenage participants not only mentioned malware as a prominent concern but also reported disproportionately high rates of being affected by it. 

 Our study also provides novel insights into teenagers' information sources for digital security compared to other adult groups. Compared to Redmiles~\etal~\cite{redmiles2016learned}, which similarly revealed age-based differences in information but did not look into teenagers in particular, our findings show teenagers' higher reliance on social media and lower reliance on sources from experts and authorities. These findings also relate to the broader literature on teenagers being on the leading edge of the social media space~\cite{Vogels_2022}.


\subsubsection{People with Migration Background\rev{s}}
\rev{While many prior studies have focused on specific migration background groups, such as undocumented immigrants and refugees, our research adopts a broader approach by employing the EU definition of migration backgrounds, which includes individuals with at least one parent born in a different country~\cite{eu-06-migration}.
With the differences in mind, we still compare our findings with these prior studies, as our definition encompasses the aforementioned more specific groups and the comparisons are based on some common grounds.
}
Our participants with migration background\rev{s} reported high rates of device usage, echoing findings that highlight the importance of ICTs within the migration process \rev{particularly for refugees}~\cite{simko2018computer, coles2019accessing, wyche2012we, brown2014reflection, Stapf2019, Kutscher2015, Gouma2020, coles2018new, lingel2014city}. 
Our participants -- most of whom \rev{might} have established lives in Germany \rev{inferred} from their education levels and language proficiency -- exhibit a similar pattern in terms of high reliance on digital devices: across all groups, they had the highest adoption rate of smartphones as well as IoT devices such as smart speakers and wearables. They also reported the most experiences with cybercrime, echoing prior work on how this group inevitably bears risks of technology use in exchange for the associated needs and benefits~\cite{guberek2018keeping,simko2018computer}.

Similar to Stapf~\cite{Stapf2019}, we also found participants with migration background\rev{s} often asked friends and family for security advice, \rev{and relied less} on official sources and consumer protection bodies, which is likely a reflection of issues in these sources not meeting their needs (\eg, the materials are hard to understand or require cultural knowledge). 

We additionally provide novel insights into this group's most salient security-related concerns -- hacker attacks and data theft -- and their perception of possible attackers -- mainly criminals and not at all family and friends.

\subsubsection{People with Low Formal Education}
In line with Redmiles~\etal~\cite{redmiles_digital_Divide}, our results indicate that we should not assume that lower education is correlated with more exposure to security threats. Redmiles~\etal\ found that people with lower educational attainment report equal or fewer incidents than more educated people~\cite{redmiles_digital_Divide}; participants with lower education in our study reported more experiences with cybercrime than the average German population, but fewer experiences than some of the more educated groups in our sample. Moreover, contrary to Redmiles~\etal~\cite{redmiles2016learned}, we did not find a digital divide in our participants' source selections either, as participants with lower education followed the same pattern as the other two adult groups by relying on online media, friends, and family as their primary sources.

We provide novel insights into this group's digital security concerns, namely hacker attacks and financial loss. Similar to people with migration background\rev{s}, they identified hackers and criminals as possible attackers more than those in their inner circle.

\subsection{Recap of Between-Group Comparisons (RQ2)}
\label{subsec:compare-between-group}

Our research joins forces with prior work that synthesizes the disjoint and sometimes contradictory digital security needs of various at-risk groups~\cite{sannon2022privacy,warford2022sok}. Moreover, our research provides novel insights into the ways in which the four at-risk groups we investigated -- older adults, teenagers, people with migration background\rev{s}, and people with low formal education -- are similar but also different in their digital security-related concerns and experiences.

\subsubsection{All Groups Trust Friends and Family}
We found that all four groups were the least concerned about people close to them (\eg, family members, friends and acquaintances, and work colleagues) posing a threat to their digital security. Friends and family were also one of the primary sources our participants used to obtain information about digital security across all groups. These findings \rev{suggest} our participants' trust in their family and friends when navigating digital security threats, and echo Warford~\etal's SoK~\cite{warford2022sok} that highlights at-risk users' reliance on social connections for advice and support. Nevertheless, such reliance comes with its own risks when people share sensitive digital resources~\cite{watson2020we} and when groups like older adults are subject to ``family surveillance'' as their family members become too paternalistic in the efforts to protect them~\cite{murthy2021individually}.

\subsubsection{Hackers Stand Out, Phishing Does Not} All groups universally viewed hackers and criminals as a threat to their digital security, although (surprisingly) only to a rather moderate extent. We can contextualize this finding in our participants' information sources with online media, print media, and TV being the major ones. \rev{Additionally}, prior work has shown how data breaches are prominently featured in security and privacy news~\cite{das2018breaking} (which can prompt concerns about hackers) and how mass media's portrayal of ``hacking'' can be inaccurate and exaggerated~\cite{gordon2010forty}, influencing users' mental models~\cite{fulton2019effect}.

Interestingly, phishing does not trigger prominent concerns for any of the four groups, despite phishing being a recurring theme in security literature and advice~\cite{redmiles2020comprehensive,reeder-17-152-simple-steps} and the prevalence of anti-phishing training programs in research and organizational settings \cite{Wash_2018,Sheng_2007, Kumaraguru_2010}. While experiences with phishing were uncommon in both our study and in the BSI survey~\cite{Digitalbarometer2022}, our participants reported falling for phishing even less frequently (between 8\% and 12\%) than respondents to the BSI survey. 

Prior work on the demographic differences in phishing susceptibility suggests that women (compared to men) and younger people (18-25 compared to those older) are more susceptible to phishing~\cite{sheng2010falls}. Adding to this body of literature, our findings suggest that even though the four groups are characterized as ``at-risk,'' they may not be at higher risk of falling for \rev{phishing} compared to the general population.
However, it is also possible that the phishing rates were under-reported in our study because phishing awareness has not reached the groups when educational efforts are distributed through the wrong channels. \rev{For example}, anti-phishing training is mostly deployed in corporate settings, while most older adults and teenagers are not employed. \rev{Another possibility is that our participants} experienced phishing but were not aware of it or equated phishing with other concepts such as unauthorized access.
The differences between our participants' concerns and the actual prevalence of phishing attacks \rev{measured in the wild}~\cite{Thomas_2017, APWG_2022} suggest that the four groups might need more education on this threat.


\subsubsection{Differences Shaped by Device Usage and Life Stages} 
Confirming prior work on the ``digital divide'' of technology use~\cite{Faverio_2022,Huxhold_2020}, our findings show that older adults are slower adopters across various digital devices, whereas teenagers and participants with migration background\rev{s} exhibit higher and more diverse device usage. Meanwhile, teenagers and participants with migration background\rev{s} also reported the most experiences with cybercrime incidents; the rates were also significantly higher than those of older adults across all types of cybercrimes except malware. While we did not conduct correlation analyses to support this, we can already see the intersection between one's device usage and exposure to cybercrimes through these numbers: with more frequent usage of various devices, the possibility of encountering security threats also rises. More device usage also lays the motivation for seeking information on how to secure different devices -- this might also explain why participants with migration background\rev{s} were the most active information seekers across all groups.

Differences in security experiences can also be shaped by one's stage of life. We observe this in multiple comparisons between teenagers and the remaining groups: participants from all adult groups reported more concerns regarding financial loss, while teenagers expressed more concerns regarding malware and password theft. This finding makes sense when contextualizing each group's concerns in their broader life stages. Teenagers tend to have fewer financial resources to manage compared to older adults, whose security and privacy concerns also often center around financial aspects~\cite{quan2020online}. Financial needs are unlikely to be the primary concern for most teenagers, whereas for people with migration background\rev{s}, especially refugees, financial needs could become competing priorities that lead them to abandon security best practices~\cite{simko2018computer}.

\subsection{Implications and Recommendations}
\label{subsec:recommendations}


Drawing on our findings, we provide education, policy, and future research recommendations for the four at-risk groups we investigated in our study and for research with at-risk user groups in general.

\subsubsection{Recommendations for Group-Specific Channels and Content} Our findings on information sources provide insights into the specific channels for reaching each group. For instance, our findings show that teenagers rely on social media for learning about digital security much more than the other groups, indicating that social media (particularly platforms like YouTube, TikTok and Instagram that have high popularity among teenagers~\cite{Vogels_2022}) can be used to reach teenagers. On the other hand, television, print media, and sources related to experts and authorities likely work better for older adults who are already using these channels for self-education.

Our findings also provide rich implications on the specific content to be prioritized in designing educational materials for these groups. For instance, for teenagers, the content could focus more on topics that they are less concerned about but report more negative experiences with, such as account compromises and data abuse; attention should also be given to how social media can fuel the learning of anti-security and anti-privacy tactics, such as those for surveilling and controlling others~\cite{wei2022anti}. For people with migration background\rev{s} and people with low formal education, it is crucial to ensure that the materials are easy to understand, in plain language, while taking into account the potential audience's diverse language skills and cultural backgrounds.

\subsubsection{Leverage Social Influence for Security Education} Prior research on social cybersecurity has shed light on how people's security decisions are subject to peer influences~\cite{das2014effect} and how certain social groups navigate security together~\cite{watson2020we,wu2022sok}. Our findings provide further empirical support for this, as friends and family were among the most utilized sources for learning about digital security across all four groups. Akin to prior work on training cybersecurity guardians in older communities~\cite{nicholson2021training}, our findings highlight the value of empowering the social circles of at-risk user groups for security education more broadly. More specific pointers can be senior centers and professional caregivers (for older adults); teachers, parents, youth clubs, sports clubs, and influencers on social media (for teenagers), organizations that serve migrants and religious groups (for people with migration background\rev{s}), organizations that serve people with lower income and local educational personal (for people with low formal education).  

Furthermore, there has been a debate on whether to embed mandatory digital security training for children and teenagers at school \cite{Sueddeutsche_2022,pencheva2020bringing}, which, if implemented, can be another source of social influence for this group. 

While social influence can be positive, it is worth mentioning how our participants rarely considered people in their inner circles (\eg, friends, family, and work colleagues) as a risk to their digital security. Such trust can be dangerous when one's close social connections become a threat vector, such as \rev{in the case of} intimate partner abuse~\cite{slupska2021threat}, elder financial abuse~\cite{acierno2010prevalence}, and parental surveillance and control~\cite{ghosh2018matter,wei2022anti}. Educational materials should highlight the possibility of interpersonal adversaries, the corresponding risks, advice on coping strategies, and links to broader resources.


\subsubsection{Protect At-Risk Groups through Policymaking}
Our findings reveal the four groups are generally more at risk of experiencing security incidents compared to the general German public (see~\autoref{results-Q6}). Thus, these groups should \rev{receive special consideration} in laws and regulations that impact one's security, privacy, and digital well-being broadly. 

One of our \rev{key} findings is that participants with migration background\rev{s} reported the highest device usage, which in turn generates more digital traces~\cite{shankar2021coordinating}; it is then unsurprising to see that they also experienced the most negative security incidents. This inevitable tradeoff this group has to make --- exchanging digital security for broader benefits associated with technology use --- indicates a failing of society and effective policies that can protect them by default and do not require them to make such tradeoffs. For example, guidance on implementing the GDPR has suggested minors and the elderly as examples of ``vulnerable persons''~\cite{ghent.gdpr.2022}, but not necessarily people with migration background\rev{s} and lower education (despite their frequent encounters with security incidents as our study suggests). On a high level, security and privacy policymaking needs to better incorporate \rev{considerations of} vulnerability by providing more explicit definitions of vulnerable data subjects as well as expanding the examples of vulnerable persons based on evidence from research~\cite{malgieri2020vulnerable}.



 \subsubsection{Future Research Directions}
 
 
Our findings add to the very limited body of literature on the security experiences of people with low formal education. However, more research could be done for a population about which so little is known. Building on our findings, we see opportunities for future research to qualitatively elicit reasons behind their concerns as well as develop and evaluate technologies that support this group's learning and self-protection.

Similarly, while our study provides first-of-its-kind empirical evidence for Warford~\etal's call of ``consider at-risk users at scale''~\cite{warford2022sok}, we believe that more can be done to shed light on the reasons behind the reported experiences for all groups. It also remains a challenge to quantitatively evaluate and compare the impact of a technology design or educational effort across multiple at-risk populations --- another direction that can be pursued by future research.
Additionally, the finding that all groups rely on friends and family as information sources with little to no concerns about interpersonal adversaries is worth exploring further. Communicating risks associated with people one knows, trusts, and delegates their digital security is a fundamentally sensitive issue. Future research could look into the specific ways of helping users develop sensible precautions and abilities to watch out for abuse without \rev{assuming} friends-and-family helpers as a security risk.



\section{Conclusion}
Our study contributes to the body of research on inclusive security and privacy by examining the digital security experiences of four at-risk groups -- older adults, teenagers, people with migration background\rev{s}, and people with low formal education -- through a large-scale ($n$$=$$1,003$) study with demographically representative samples for each group. 
Since demographically representative samples for these groups can not or not easily be obtained from online panels, we used computer-assisted telephone interviews (CATIs) to investigate participants' device usage, security concerns, prior cybercrime incidents, perceptions of potential attackers, and information sources for security (RQ1), as well as the differences and similarities between the four groups (RQ2).
Our results show that participants with low formal education do not have distinctive patterns compared to participants with migration background\rev{s}, but \rev{exhibit} significant differences compared to teenagers and older adults. 
Teenagers and participants with migration background\rev{s} had higher and more diverse device usage while reporting the most experiences with cybercrime. 
Conversely, older adults indicated lower device usage, were less affected by cybercrime, and had lower risk perceptions regarding possible attackers. The adult sample groups relied more on traditional information sources, whereas teenagers mainly obtained information about digital security from social media. All groups similarly identified friends and family and online media as their most used information sources and did not regard their social circles as possible attackers. 
Our research lays the foundation for more cross-group comparisons and syntheses of at-risk users' diverse experiences. 
Our findings also help identify specific educational approaches, policy recommendations, and directions for future work.




\section*{Acknowledgments}
We would like to thank Leonie Schaewitz, Carina Wiesen, and Jennifer Friedauer for their support, as well as all participants in our study and our CATI provider.  
This work was supported by the Deutsche Forschungsgemeinschaft (DFG, German Research Foundation) under Germany’s Excellence Strategy -- EXC 2092 CASA -- 390781972 and by the federal state of NRW, Germany through the PhD School ``SecHuman -- Security for Humans in Cyberspace''.

\bibliographystyle{IEEEtran}
\bibliography{CATIBib}

\appendices
\section{Complete CATI Questionnaire}
\label{appendix:interview}

\definecolor{headlines}{HTML}{58429B}
\definecolor{info}{HTML}{00AEB3}
\definecolor{description}{HTML}{103778}

\renewcommand{\labelitemi}{$\bullet$}
\newcommand{\breakparagraph}[1]{\paragraph*{#1}\mbox{}\\}
\small

\noindent\textbf{{\color{headlines}Demographics}}

    \noindent \textbf{Q\_Age:} \textbf{How old are you?} \textit{{\color{info}}}
        \begin{itemize}
            \item {\color{description}Items}: \emph{14-17; 18-35; 36-50; 51-65; 66-69; 70-79; 80+}
        \end{itemize}
    \textbf{Q\_Gender:} \textbf{What is your gender?} \textit{{\color{info}}}
        \begin{itemize}
        \item {\color{description}Items}: \emph{Female; Male; Non-binary; Describe yourself: \textit{{\color{info}[free response]}}; Prefer not to answer.}
        \end{itemize}
        \textbf{Q\_State:} \textbf{In which state do you live?} \textit{{\color{info}}}
        \begin{itemize}
            \item {\color{description}Items}: \emph{Baden-Württemberg; Bavaria; Berlin; Brandenburg; Bremen; Hamburg; Hesse; Lower Saxony; Mecklenburg-Western Pomerania; North Rhine-Westphalia; Rhineland-Palatinate; Saarland; Saxony; Saxony-Anhalt; Schleswig-Holstein; Thuringia}
        \end{itemize}
    \textbf{Q\_Nationality:} \textbf{Were you or at least one part of your parents born with a foreign nationality?} \textit{{\color{info}}}
        \begin{itemize}
            \item {\color{description}Items}: \emph{Yes; No}
        \end{itemize}
    \textbf{Q\_Education:} \textbf{What is your highest level of education?} \textit{{\color{info}}}
        \begin{itemize}
            \item {\color{description}Items}: \emph{No school leaving certificate; Secondary school (primary school) or equivalent leaving certificate; High school (O level) or equivalent leaving certificate; A level, vocational high school / general or university entrance qualification; Occupational or vocational training / apprenticeship; Completion of a technical college or administrative or professional academy; Bachelor’s degree; Diploma university course or masters (including: teaching position, state examination, Master’s course, artistic or comparable courses of study); PhD/doctorate; Prefer not to answer.}
        \end{itemize}

\vspace{1em}
\noindent\textbf{{\color{headlines}Internet Usage}}\\
First, I would like to ask you some questions about your internet usage.

\begin{enumerate}
    \item[\textbf{Q1}] \textbf{I'm going to read through a list of devices. Please tell me for each device whether you use it in your daily life or not.} \textit{{\color{info}[multiple choice]}}
    \begin{itemize}
        \item {\color{description}Items}: \emph{Smartphones; Static PCs / desktop PCs; Laptops; Tablets; Voice assistants or smart speakers (\eg, Alexa, Amazon Echo); Wearables (\eg, fitness trackers or smartwatches)}
    \end{itemize}
        \item[\textbf{Q2}] \textbf{How often do you use the internet for the following purposes? I'm going to read you a list of purposes and you indicate how often you're using the internet for these purposes.} 
    \begin{itemize}
        \item {\color{description}Items}: \emph{Online shopping; Ordering services (e.g. booking travel, ordering food); Selling goods or services (e.g. via eBay); Researching information and forming opinions (e.g. reading online newspapers); Uploading and sharing personal content you have created yourself (texts, images, photos, videos) ; Expressing opinions (e.g. posts on social media); Online banking; Communication (e.g. via email and chat); Entertainment (e.g. streaming films, online games); Official transactions (e.g. ordering an identity card); Health services (e.g. electronic patient record, virtual doctor appointment); Map services (e.g. Google Maps or navigation services); Data storage via cloud services}
        \item {\color{description}Answer Options}: \emph{1-Never; 2-Less than once a month; 3-At least once a month; 4-At least once a week, 5-every day; Prefer not to answer.}
    \end{itemize}
    \item[\textbf{Q3}] \textbf{Next, it's about how you communicate digitally. I'm going to read through a list of communication channels and you tell me in each case how often you use the following communication channels}. 
    \begin{itemize}
        \item {\color{description}Items}: \emph{Email; Calling via stationary phone; Calling with your smartphone or cell phone; SMS; Messenger services (such as WhatsApp or Signal); Social media (such as Facebook or Instagram); Online forums and communities; Video calls (for example via Skype, Zoom, or Microsoft Teams}
        \item {\color{description}Answer Options}: \emph{Never; Less than once a month; At least once a month; At least once a week, Daily; Prefer not to answer.}
    \end{itemize}
    \item[\textbf{Q4}] \textbf{Reflecting on the topic of digital security: Is there anything you're concerned about?} Please name anything that comes to your mind spontaneously \textit{{\color{info}[free response]}}
        \item[\textbf{Q5}] \textbf{How familiar are you with the following terms?} \textit{{\color{info}}}
    \begin{itemize}
        \item {\color{description}Items}: \emph{Malicious software (for example a computer virus); Ransomware; Phishing; Spear phishing; Two-factor authentication (2FA); Biometric authentication methods; Identity theft; Data leakage or data theft; HTTPS; Hard disk encryption; End-to-end encryption; Transport encryption; Browser; Private browser mode (respectively incognito mode); IP address; URL; Virtual Private Network (VPN); Tor network; ad blocker; Love scam (respectively online love fraud); Spam; Cloud)}
        \item {\color{description}Answer Options}: \emph{I have never heard of this; I have heard about it, but I don't know how it works; I know what it is and how it works; Prefer not to answer.}
    \end{itemize}
        \item[\textbf{Q6}] \textbf{The next question is about your experiences with cybercrime. Have you been affected to cybercrime yourself? I'm going to read through a list of items and ask you to tell me whether you have ever been affected by them or not.} \textit{{\color{info}[multiple choice]}}
    \begin{itemize}
        \item {\color{description}Items}: \emph{Malware (such as viruses or Trojans); Phishing, \ie, spying out of confidential data; Ransomware or cryptoviral extortion; Cyberbullying; Online shopping fraud; Foreign access to your online account; Cyberstalking; Victims of data misuse, \ie, the disclosure or sale of personal data (\eg, your telephone number, address, or bank details); Love scam (\ie, love fraud on the internet)} 
        \item {\color{description}Answer Options}: \emph{Yes; No; Prefer not to answer.}
    \end{itemize}
        \item[\textbf{Q7}] \textbf{Do you inform yourself about the topic of \emph{digital security}?} \textit{{\color{info}}}
          \begin{itemize}
        \item \emph{Yes; No}
    \end{itemize}
    \item[\textbf{Q8}] \textit{{\color{info}[If ``Yes'' in Q7]}} \textbf{The next question is about where you seek information on the topic of digital security. I'll read through the list once again, but related to information sources and you tell me if you're use this respective source to inform yourself on the topic of digital security} \textit{{\color{info}[multiple choice]}}
    \begin{itemize}
        \item {\color{description}Items}: \emph{Print media; Social media (such as Facebook or Instagram); Radio and/or podcasts; Television; Friends and/or acquaintances and/or family; IT security experts; Consumer advice centers and authorities} 
        \item {\color{description}Answer Options}: \emph{Yes; No; Prefer not to answer.}
    \end{itemize}
    \item[\textbf{Q9}] You're almost done, there are only a few questions left. \textbf{Up next is what data you would like to protect and who you would like to protect your data from. I will read out types of data and ask you to tell me in each case how important it is to you to protect this data on the Internet, for example from outside access and theft.} 
    \begin{itemize}
        \item {\color{description}Items}: \emph{Your full name; your address or home address; your home telephone number; your contacts; your private photos; message histories (for example, chat and emails); Location and movement histories (for example, GPS data from your jogging route); Amount of salary or earnings; Identification documents (such as, ID card and driver's license); Insurance documents; Delivery bills and invoices; IBAN and BIC, or amount data; Health data; Biometric data (such as fingerprints); Passwords} 
        \item {\color{description}Answer Options}: \emph{1-Not important; 2-A little important; 3-Moderately important; 4-Quite-a-bit important; 5-Very important}
    \end{itemize}
\end{enumerate}
\begin{enumerate}
    \item[\textbf{Q10}] \textbf{I'm going to read through a yet another list about groups of people. For each of these groups of people, please tell me how likely you think it is that this group people poses a risk to your digital security  -- for example, unauthorized access to your personal data, stalk you online or restrict your access to digital services.} 
    \begin{itemize}
        \item {\color{description}Items}: \emph{Family members; Friends and acquaintances; Work colleagues; Officials from Germany, such as police, secret services and the government; Officials from other countries, such as police, secret services and the government; Private sector companies; Criminals who want to get rich from your data; Hackers who gain unauthorized access to data and devices, for fun.}
        \item {\color{description}Answer Options}: \emph{ 1-not likely; 2-a little likely; 3-moderately likely; 4-quite a bit likely; 5-very likely.}
    \end{itemize}

\end{enumerate}
\onecolumn
\newpage
\section{Codebook for Question 4}
\label{appendix:codebookCATI}

\begin{table*}[ht]
    \caption{Full codebook for Q4 (``Reflecting on the topic of digital security: Is there anything you are concerned about?'') and assignment frequencies for each of the four CATI subgroups (teenagers, older adults, participants with migration backgrounds, and participants with low formal education).}
    \label{appendix:codebookcomplete}
    \centering
    \footnotesize
    \rowcolors{2}{white}{gray!10}
    \scalebox{1.0}{
    \begin{tabular}{lcccS[table-format=3]S[table-format=2]S[table-format=3]S[table-format=2]S[table-format=3]S[table-format=3]}
        \toprule
        \rowcolor{white}
        \multicolumn{4}{l}{\textbf{Code}} & \multicolumn{5}{c}{\textbf{\textbf{CATI}}} \\ 
        
        \rowcolor{white}
        
        \multicolumn{4}{c}{} & \multicolumn{1}{c}{Teenagers} & \multicolumn{1}{c}{O. Adults} & \multicolumn{1}{c}{Migration B.} & \multicolumn{1}{c}{Low Education} & \multicolumn{1}{c}{Complete} \\
        \multicolumn{4}{c}{} & \multicolumn{1}{c}{$n = 96$} & \multicolumn{1}{c}{$n = 85$} & \multicolumn{1}{c}{$n = 123$} & \multicolumn{1}{c}{$n = 123$} & \multicolumn{1}{c}{$n = 428$} \\
        \midrule

            \multicolumn{4}{l}{
            \textbf{Active Attack}} & & & &  &
           \\

            {} & \multicolumn{3}{l}{
            \emph{Unauthorized access to (your) devices}} & 9 & 2 & 5 & 8 &
            24 \\

            {} & \multicolumn{3}{l}{
            \emph{Financial loss}} & 7 & 18 & 19 & 15 &
            59 \\

            {} & \multicolumn{3}{l}{
            \emph{Hacker attack}} & 23 & 14 & 23 & 23 &
            83 \\

            {} & \multicolumn{3}{l}{
            \emph{Data theft (unnoticed)}} & 13 & 5 & 23 & 13 &
            54 \\

            {} & \multicolumn{3}{l}{
            \emph{Cyberbullying or Cyberstalking}} & 7 & \multicolumn{1}{c}{~-} & 1 & 1 &
            9 \\

            {} & \multicolumn{3}{l}{
            \emph{Fraud}} & 2 & 7 & 7 & 12 &
            28 \\

            {} & \multicolumn{3}{l}{
            \emph{Malware}} & 20 & 3 & 9 & 9 &
            41 \\

            {} & \multicolumn{3}{l}{
            \emph{Password theft}} & 16 & 1 & 7 & 6 &
            30 \\

            {} & \multicolumn{3}{l}{
            \emph{Phishing}} & 7 & 3 & 8 & 5 &
            23 \\

            {} & \multicolumn{3}{l}{
            \emph{Involuntary publication of personal data}} & 5 & \multicolumn{1}{c}{~-} & 5 & \multicolumn{1}{c}{~-} &
            10 \\

            {} & \multicolumn{3}{l}{
            \emph{Fake accounts}} & 1 & \multicolumn{1}{c}{~-} & 1 & 2 &
            4 \\

            {} & \multicolumn{3}{l}{
            \emph{Data misuse}} & \multicolumn{1}{c}{\quad-} & 2 & 3 & 1 &
            6 \\

            {} & \multicolumn{3}{l}{
            \emph{Criminals}} & 1 & 2 & 5 & 4 &
            12 \\

            {} & \multicolumn{3}{l}{
            \emph{Identity theft}} & 1 & 5 & 8 & 1 &
            15 \\

            \multicolumn{4}{l}{\parbox[t]{3.5cm}{
            \textbf{Tracking}}} & & & &  &
            \\

            {} & \multicolumn{3}{l}{
            \emph{Data collection, aggregation, and use}} & 4 & 5 & 11 & 6 &
            26 \\

            {} & \multicolumn{3}{l}{
            \emph{Unintentional data disclosure}} & 3 & 2 & 8 & 4 &
            17 \\

            {} & \multicolumn{3}{l}{
            \emph{Profiling}} & \multicolumn{1}{c}{\quad-} & \multicolumn{1}{c}{~-} & \multicolumn{1}{c}{\quad-} & 1 &
            1 \\

            {} & \multicolumn{3}{l}{
            \emph{Cookies}} & 1 & 1 & 2 & 4 & 
            8 \\

            {} & \multicolumn{3}{l}{
            \emph{Personalized advertising}} & 3 & 4 & 1 & 4 &
            12 \\

            {} & \multicolumn{3}{l}{
            \emph{Forced disclosure of personal data}} & 1 & 1 & 2 & 2 &
            6 \\

            \multicolumn{4}{l}{\parbox[t]{3.5cm}{
            \textbf{Passive Attack}}} & &  & & &
             \\

            {} & \multicolumn{3}{l}{
            \emph{Eavesdropping}} & 3 & 2 & 5 & 2 & 
            12 \\

            {} & \multicolumn{3}{l}{
            \emph{Data spying}} & 2 & 3 & 1 & 3 &
            9 \\

            {} & \multicolumn{3}{l}{
            \emph{Lack of data protection}} & \multicolumn{1}{c}{\quad-} & 3 & 10 & 9 & 
            22 \\

            {} & \multicolumn{3}{l}{
            \emph{Surveillance}} & 1 & 9 & 5 & 10 &
            25 \\

            \multicolumn{4}{l}{\parbox[t]{3.5cm}{
            \textbf{General Concerns}}} &  & &  &  & 
             \\

            {} & \multicolumn{3}{l}{
            \emph{Internet security}} & 2 & 6 & 7 & 10 &
            25 \\

            {} & \multicolumn{3}{l}{
            \emph{Data loss}} & 3 & \multicolumn{1}{c}{~-} & 5 & 1 &
            9 \\

            {} & \multicolumn{3}{l}{
            \emph{Data protection}} & 7 & 2 & 7 & 5 &
            21 \\

            \multicolumn{4}{l}{\parbox[t]{3.5cm}{
            \textbf{Loss of Control}}} &  & & &  &
             \\

            {} & \multicolumn{3}{l}{
            \emph{Lack of transparency}} & 1 & 2 & 2 & 3 &
            8 \\

            {} & \multicolumn{3}{l}{
            \emph{Dependency on digital media}} & \multicolumn{1}{c}{\quad-} & \multicolumn{1}{c}{~-} & 1 & 1 & 
            2 \\

            {} & \multicolumn{3}{l}{
            \emph{Lack of information (about fraud schemes)}} & 1 & 1 & 2 & \multicolumn{1}{c}{~-} & 
            4 \\

            {} & \multicolumn{3}{l}{
            \emph{Life shifts to the virtual world}} & \multicolumn{1}{c}{\quad-} & 1 & \multicolumn{1}{c}{\quad-} & \multicolumn{1}{c}{~-} &
            1 \\

            {} & \multicolumn{3}{l}{
            \emph{No digital forgetting}} & 1 & 1 & \multicolumn{1}{c}{\quad-} & 4 &
            6 \\

            {} & \multicolumn{3}{l}{
            \emph{Lack of protection and education for children}} & \multicolumn{1}{c}{\quad-} & 1 & 3 & 1 &
            5 \\

            {} & \multicolumn{3}{l}{
            \emph{Speed of digitalization}} & 1 & 1 & 2 & \multicolumn{1}{c}{~-} &
            4 \\

            {} & \multicolumn{3}{l}{
            \emph{Internet as a lawless space}} & \multicolumn{1}{c}{\quad-} & \multicolumn{1}{c}{~-} & \multicolumn{1}{c}{\quad-} & 2 &
            2 \\

            \multicolumn{4}{l}{\parbox[t]{3.5cm}{
            \textbf{Non-targeted Attack}}} & & \multicolumn{1}{c}{~-} &  &  &
             \\

            {} & \multicolumn{3}{l}{
            \emph{Spam}} & 2 & \multicolumn{1}{c}{~-} & 1 & 3 &
            6 \\

            \multicolumn{4}{l}{\parbox[t]{3.5cm}{
            \textbf{No Concerns}}} & 4 & 3 & 1 & 4 &
            12 \\

            \multicolumn{4}{l}{\parbox[t]{3.5cm}{
            \textbf{No Codes Possible}}} & 8 & 8 & 4 & 8 &
            28 \\
        \bottomrule
    \end{tabular}}
\end{table*}

\end{document}